\documentclass[aps,pra,twocolumn,floatfix,10pt]{revtex4} 
\usepackage{graphicx,amsmath,amssymb,ulem}
\usepackage{textcomp}
\usepackage{hyperref}
\usepackage{color}
\usepackage{epstopdf}
\usepackage{nicefrac}
\usepackage[caption=false]{subfig}
\usepackage{gensymb}
\usepackage{placeins}
\usepackage{multirow}
\usepackage[flushleft]{threeparttable}

\def\K#1{$^{#1}$K}
\def\2K{$^{39}$K$_2$}

\def\Xstate{$X^1\Sigma^+_g$~}
\def\astate{$a^3\Sigma^+_u$~}
\def\wn{cm$^{-1}$}

\def\ket#1{\mathinner{|{#1}\rangle}}

\begin{document}

\title{Beyond Born-Oppenheimer approximation in ultracold atomic collisions}

\author{Eberhard~Tiemann}
\email{tiemann@iqo.uni-hannover.de}
\author{\\Philipp~Gersema}
\author{Kai~K.~Voges}
\author{Torsten~Hartmann}
\author{Alessandro~Zenesini}
\author{Silke~Ospelkaus}
\email{silke.ospelkaus@iqo.uni-hannover.de}

\affiliation{Institut f\"ur Quantenoptik, Leibniz Universit\"at Hannover, 30167~Hannover, Germany}

\date{\today}

\begin{abstract}
 We report on deviations beyond the Born-Oppenheimer approximation in the potassium inter-atomic potentials. Identifying three up-to-now unknown $d$-wave Feshbach resonances, we significantly improve the understanding of the $^{39}$K inter-atomic potentials. Combining these observations with the most recent data on known inter- and intra-isotope Feshbach resonances, we show that Born-Oppenheimer corrections can be determined from atomic collisional properties alone and that significant differences between the homo- and heteronuclear case appear.
\end{abstract}
\maketitle

\section{Introduction}

In quantum chemistry and molecular physics, the assumption that the electronic and nuclear motions can be separately treated is well justified by the three orders of magnitude separating the proton and the electron mass. The nuclei are considered as fixed objects at relative distance $R$ when solving the eigenvalue problem of the electron motion resulting in a $R$-dependent electronic energy which is taken as the potential for the nuclear motion. This approximation leads to large simplifications when solving the Schr\"{o}dinger equation for molecules and is named \textit{Born-Oppenheimer Approximation} (BOA)\cite{BornOpp}. The BOA is extremely powerful in matching theoretical predictions and spectroscopic results, in particular concerning the understanding of diatomic molecules. One major aspect within the BOA is that the same interatomic potential (BO-potential) is used for different isotopes by simply rescaling the nuclear motion according to the reduced molecular mass. Deviations from this assumption lead to perturbative corrections to the BOA on the order of the electron-to-proton mass ratio. The isotopic dependence of corrections has been discussed in many papers, see for example \cite{HERMAN1966305,Watson1980,Bunker1977}. These deviations from the BO-potential approach correspond to shifts in energy levels on the order of $\Delta E/E\approx 10^{-4}$ or less and they have been observed in spectroscopy experiments like \cite{Che1061,Knoeckel2004,PhysRevA.78.012503} and in the dissociation energy of different isotope combinations of hydrogen diatomic molecules \cite{Farnik2016}.\\
Effects of the corrections to the BOA are much weaker at the collisional threshold of atom pairs as the long range behaviour of the inter-atomic potential is weakly affected by short range variations. Recent developments in molecule cooling and molecule association from ultracold atoms have considerably increased the experimental resolution giving access to study effects beyond the BOA. Corrections to the triplet and singlet scattering length are indeed predicted to be on the order of few tenths of the Bohr radius $a_0$ \cite{PhysRevA.78.012503} and are practically undetectable. However, Feshbach resonances are an effective passkey as they carry  important information of the short range potential to the atomic threshold \cite{RevModPhys.82.1225}. A particular interesting case is given by the collisional properties of ultracold potassium atoms. Potassium features two stable bosonic isotopes (\K{39} and \K{41}) and a very long living fermionic one (\K{40}). All these isotopes have been cooled to quantum degeneracy both in single and in two isotope experiments and collisional data for five (\K{39}-\K{39}, \K{40}-\K{40}, \K{41}-\K{41}, \K{39}-\K{41}, \K{40}-\K{41}) of the six possible combinations are available in literature \cite{PhysRevLett.90.053201, D_Errico_2007, PhysRevA.84.011601, PhysRevA.94.033408, PhysRevA.98.062712}. The comparison of Feshbach resonance positions for different isotope combinations is a promising way to reveal corrections to the BOA. First hints of such corrections were obtained by Falke \textit{et al.} \cite{PhysRevA.78.012503} studying the two cases \K{39}-\K{39}, and \K{40}-\K{40} available at that time.
In this paper we present the experimental observation of up-to-now unmeasured d-wave Feshbach resonances of \K{39} and how this allows to improve the knowledge on \2K. We combine this result with the published literature on potassium Feshbach resonances and we determine corrections to the BOA from collisional data alone.\\\newline
The paper is structured as follows. In Section \ref{Theory} we explain how to reveal effects from beyond BOA-corrections  in atomic collisional properties. In Section \ref{dwave} we present the observation of three new Feshbach resonances for \K{39}, which enhances the knowledge on \2K to use this dimer as reference for the full isotope analysis. In Section \ref{fit} we quantify the corrections to the BOA thanks to a multi-parameter fit of the new and already known Feshbach resonance positions.

\section{Theory aspects}
\label{Theory}
To treat the collision of an atom pair of alkali atoms at low kinetic energy we set up the Hamiltonian of the coupled system of the two lowest molecular states \Xstate and $a^3\Sigma^+_u$, because the product state of two ground state atoms is generally a mixture of singlet and triplet states. The appropriate Hamiltonian is presented in many papers, e.g. \cite{RevModPhysFesh,PhysRevA.78.012503} and will not be repeated here. It contains the hyperfine interaction, also responsible for the singlet-triplet mixing, the atomic Zeeman interaction and the effective spin-spin interaction of the two atoms in their doublet states. The nuclear motion is governed by the molecular potentials of the two interacting molecular states.\\
The potential functions within the Born-Oppenheimer approximation (BO-potentials) are represented in analytic form as described in detail in \cite{PhysRevA.78.012503} in three $R$-sections divided by an inner $R_\mathrm{in}$ and outer radius $R_\mathrm{out}$: \\\newline
In the intermediate range around the minimum it is described by a finite power expansion
\begin{equation}
\label{uanal}
U_{\mathrm {IR}}(R)=\sum_{i=0}^{n}a_i\,\xi(R)^i
\end{equation}
with a nonlinear variable function $\xi$ of internuclear separation $R$:
\begin{equation}
\label{xv}
\xi(R)=\frac{R - R_m}{R + b\,R_m}.
\end{equation}
In Eq.\,\ref{uanal} the \{$a_i$\} are fitting parameters and $b$  and $R_m$
are chosen such that only few parameters $a_i$ are needed for describing the steep slope at the short internuclear separation side and the smaller slope at the large $R$ side by the analytic form of Eq.\,\ref{uanal}. 
$R_m$ is normally close to the value of the equilibrium separation.\\\newline
The potential is extrapolated for $R < R_\mathrm{in}$ with:
\begin{equation}
\label{rep}
  U_{\mathrm {SR}}(R)= A + B/R^{N_s}
\end{equation}
\noindent by adjusting the $A$ and $B$ parameters to get a continuous transition at $R_\mathrm{in}$; 
the final fit uses $N_s$ equal to 12 and 6 for \Xstate~and \astate\ states, respectively, as adequate exponents.\\\newline
For large internuclear distances ($R > R_\mathrm{out}$)
we adopt the standard long range form of molecular potentials:
\begin{equation}
\label{lrexp}
  U_{\mathrm {LR}}(R)=U_{\infty}-\frac{C_6}{R^6}-\frac{C_8}{R^8}-\frac{C_{10}}{R^{10}}\pm E_{\mathrm{exch}}(R)
\end{equation}
\noindent where the exchange contribution is given by
\begin{equation}
\label{exch}
E_{\mathrm{exch}}(R)=A_{\mathrm{ex}} R^\gamma \exp(-\beta R) 
\end{equation}
and $U_{\infty}$ is set to zero which fixes the energy reference of the total potential scheme.\\\newline
The BO-potentials are extended by correction functions $U_\mathrm{ad}(R)$, which make the full potentials mass dependent. These correction functions \cite{Watson1980,Bunker1977,vanVleck1936} contain matrix elements of the nuclear momentum operators over the electronic wavefunctions of the considered electronic state and other ones with $\Delta \Omega =0$,  where $\Omega$ is the projection of the total electronic angular momentum onto the molecular axis. $U_\mathrm{ad}(R)$ is the so called adiabatic correction to the BO-potential function, and it contains the interaction of the considered electronic state with all states according the selection rule $\Delta \Omega=0$ by the nuclear vibrational motion. We do not include non-adiabatic correction with the selection rule $\Delta \Omega=\pm 1$ because it will be negligibly small for collisions with low partial waves as $s$, $p$ or $d$.\\
Watson \cite{Watson1980} shows that in the lowest order the mass dependence of these corrections for a molecule $AB$ will be of the form $U_A(R)\frac{m_e}{M_A}+U_B(R)\frac{m_e}{M_B}$, where $M_{A(B)}$ is the atomic mass of atom $A(B)$ and $m_e$ the electron mass. For true heteronuclear molecules the coefficients $U_A(R)$ and $U_B(R)$ will be different, for homonuclear cases in the electronic system such as K$_2$ both coefficient will be equal and thus the isotope dependence of the correction function will be inversely proportional to the reduced mass $\mu$ of the molecule. van Vleck \cite{vanVleck1936} considered the mass dependence of the heteronuclear cases in the hydrogen-deuterium (HD) molecule and found that the corrections should be extended by a term $(M_A-M_B)^2/(M_A+M_B)^2$. Thus in our case with the nuclei like \K{39}-\K{41}, the representation of the correction functions should read:
\begin{equation}
\label{ad}
U_\mathrm{ad}(R)=U_\mathrm{gen}(R)\frac{m_e}{\mu}+U_\mathrm{asym}(R)m_e\left(\frac{M_A-M_B}{M_A+M_B}\right)^2,
\end{equation}
where $\mu$ is the reduced mass for the molecular rovibrational motion. $U_\mathrm{gen}(R)$ and $U_\mathrm{asym}(R)$ are functions of the internuclear separation $R$. The subscripts refer to the general and asymmetric contributions.

\section{\textit{d\,}-wave Feshbach resonances in \K{39}}     
\label{dwave}

In our setup, ultracold samples of \K{39} atoms are prepared by sympathetic cooling in a bath of evaporatively cooled Na atoms as explained in \cite{SchulzeBEC2018,Hartmann2019}.  
Compared to the experimental sequence used in our previous works, the mixture here is heavily unbalanced towards \K{39} and the Na atom number is practically negligible. During evaporation in a crossed optical dipole trap, the \K{39} atoms are initially in the $|f=1,m_f=-1\rangle$ state and are transferred to the target $|f=2,m_f=-2\rangle$ state by rapid adiabatic passage. $f$ is the total angular momentum of the atom and $m_f$ its projection. The transfer is based on a 1\,ms radio frequency sweep performed at an external magnetic field of about 199\,G. At this magnetic field losses are small both in the initial and final state and during the transfer. The sample contains up to $3\times10^5$ atoms at 650\,nK in a trap with an average frequency of $2\pi\times114(5)$\,Hz.\\
\begin{figure}[h]
	\includegraphics[width=1\columnwidth]{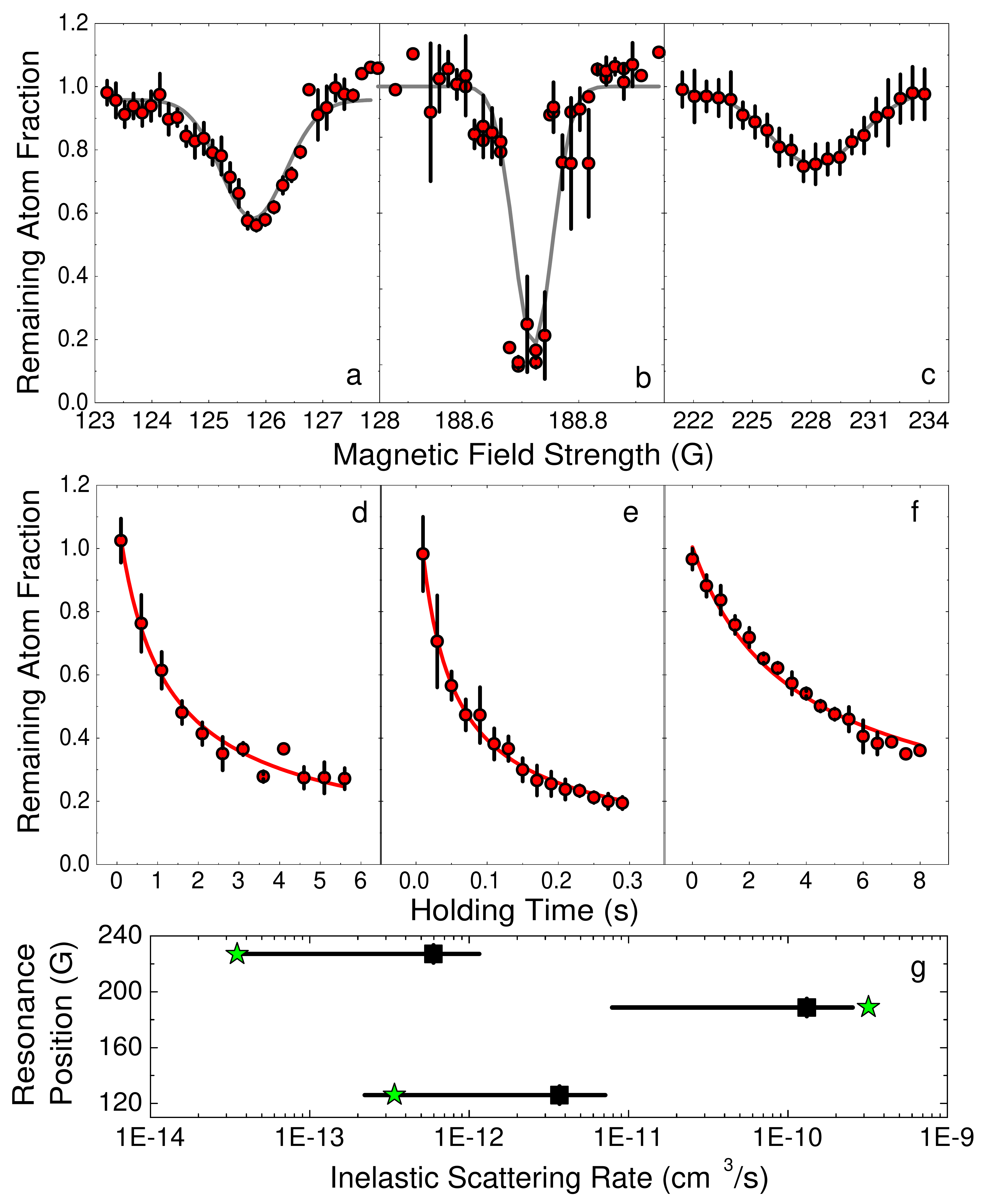}
	\caption{ \K{39} $d$-wave Feshbach resonances in the hyperfine state $|f=2,m_f=-2\rangle$. (a, b, c) The remaining fraction of \K{39} atoms is measured for different values of the magnetic field. The holding times for the three measurements are 1\,s, 470\,ms and 4\,s, respectively. Gray lines are the phenomenological Gaussian fit curves. Error bars are the standard deviation of different experimental runs. (d, e, f) Remaining atom fraction as a function of time for the $d$-wave Feshbach resonances in panel (a), (b) and (c), respectively. The measurements are performed at the magnetic fields where losses are maximal. The lines are fits according to the two-body loss rate equation; see text. (g) Comparison between the measured (black squares) and predicted (green stars) inelastic loss rates. For the uncertainties on the loss rate see text.}
	\label{Losses}
\end{figure}
To locate the $d$-wave resonances, we observe the atom number decreased due to inelastic two-body losses in the proximity of the Feshbach resonance. We ramp the magnetic field strength in 10\,ms to the target value. After a fixed holding time, chosen to not lead to complete depletion of the atoms at resonance, the magnetic field is ramped back to the magnetic field strength where high-field absorption imaging of the remaining atoms is performed \cite{Hartmann2019}. Figure \ref{Losses} shows the remaining atom fraction at different values of the magnetic field strength in the vicinity of the predicted $d$-wave Feshbach resonances. By fitting the loss data with phenomenological Gaussian curves we obtain the following three resonance positions: 125.94(14)\,G, 188.72(5)\,G and 227.71(60)\,G. The predicted width of the resonance at 188.72\,G is far below our magnetic field stability of about 30\,mG and leads to experimental points not following a Gaussian profile, compare Fig.\,\ref{Losses}(b).\\
We also measure the remaining atom number at the resonance positions for variable holding time. The data are shown in Fig.\,\ref{Losses}(d), (e) and (f) for the 125.94\,G, 188.72\,G and 227.71\,G resonance, respectively. The inelastic loss rate coefficients are obtained from a fit to the data according to the two-body loss differential equation including the effects of anti-evaporation heating \cite{PhysRevLett.91.123201, Gregory2019p} and background lifetime (about 17\,s). The loss rate coefficients are summarized in Fig.\,\ref{Losses}(g) and confirm the expected large difference between fast (Fig.\,\ref{Losses}(e)) and slow (Fig.\,\ref{Losses}(d) and (f)) losses despite the same $d$-wave character of the resonances. The values are in good agreement with theoretical predictions using the results of Sec.\,\ref{fit} within the error bars, which include statistical uncertainties and the uncertainties on the calibration of temperature, trap frequency and atom number. The values for the 125.94\,G (Fig.\,\ref{Losses} (a,d)) and 227.71\,G (Fig.\,\ref{Losses} (c,f)) resonances are larger than predicted, probably because of other loss contributions not included in the fit. The measured value for the 188.7\,G (Fig.\,\ref{Losses} (b,e)) resonance is instead smaller than expected as the narrow resonance width and the magnetic field jitter do not allow to remain exactly at resonance.

\section{Analysis}
\label{fit}
We start our analysis from potential functions of the two lowest electronic states \Xstate and \astate\  derived from spectroscopic observation which have been described in detail in \cite{PhysRevA.78.012503}. We refit the spectroscopic data with a smaller set of potential parameters to reduce  the risk of obtaining unphysical tiny oscillatory behavior of the potential function. In \cite{PhysRevA.78.012503}  state \Xstate was described by 31 parameters, now 22 are sufficient. For state  \astate we use  14  parameters compared to 22 in the previous work. The resulting potentials form the starting point for a fit of 49 Feshbach resonances and a comparison of the experimentally determined  Feshbach resonance positions with the ones resulting from the coupled channel calculation. We identify the Feshbach resonance position by the maximum scattering rate coefficient at the kinetic energy given by the experimental conditions.
\begin{table*}
\fontsize{8pt}{13pt}\selectfont
\caption{Feshbach resonances and their theoretical modeling. Columns o-c give the difference of observed field value (obs, and experimental uncertainty (unc)) to calculated one (not listed). The number indicates the model of evaluation as in Tab.\,\ref{Sigma}: (1) pure BO-potentials, (2) adding BO-correction $BO_\mathrm{gen}$ for the general case, (3) adding BO-corrections  $BO_\mathrm{gen}$ and $BO_\mathrm{asym}$ for the homo- and heteronuclear case. }
\label{tab:FR}
\begin{tabular*}{1.7\columnwidth} {@{\extracolsep{\fill}}cccccccc|ccc|c}
\hline
isotope& $M$& atom pair& $l_\mathrm{min}$&$l_\mathrm{max}$&type& obs [G]& unc [G]&o-c(1) [G]  &o-c(2) [G]& o-c(3) [G] &ref. \\
\hline
39/39&  2 &  $\ket{1,1}$ $\ket{1,1}$ & 0  & 0&el&  403.4  &  0.7  &  0.958 &  0.876  &  0.849  & \cite{D_Errico_2007}\\
     &  2 & $\ket{1,1}$ $\ket{1,1}$ & 0  & 0&el&  752.3  &  0.1  &  0.104 &  0.024  &  0.039  & \\
     &  2 & $\ket{1,1}$ $\ket{1,1}$ & 0  & 0&el&   25.85 &  0.1  &  0.016 & -0.033  & -0.024  & \\
     &  0 & $\ket{1,0}$ $\ket{1,0}$ & 0  & 2\footnote{This $s$-wave resonance is influenced by a $d$-wave resonance at 59.9 G.}&el&   59.3  &  0.6  &  0.400 &  0.315  &  0.320  & \\
     &  0 & $\ket{1,0}$ $\ket{1,0}$ & 0  & 0&el&   66.0  &  0.9  &  0.491 &  0.421  &  0.430  & \\
     & -2 & $\ket{1,\text{-}1}$ $\ket{1,\text{-}1}$ & 0  & 0&el&   32.6  &  1.5  & -1.006 & -0.979  & -0.982  & \\
     & -2 & $\ket{1,\text{-}1}$ $\ket{1,\text{-}1}$ & 0  & 0&el&  162.8  &  0.9  &  0.503 &  0.460  &  0.467  & \\
     & -2 & $\ket{1,\text{-}1}$ $\ket{1,\text{-}1}$ & 0  & 0&el&  562.2  &  1.5  &  1.383 &  1.301  &  1.274  & \\
     & -2 & $\ket{1,\text{-}1}$ $\ket{1,\text{-}1}$ & 0  & 2&in&  395.1  & 1.0   & -0.020 & -0.122  & -0.160  & \cite{Fouche2019}\\
     & -1 & $\ket{1,0}$ $\ket{1,\text{-}1}$ & 0  & 0&el&  113.76 & 0.1   &  0.062 & -0.002  &  0.007  & \cite{PhysRevA.98.062712}\\
     & -4 & $\ket{2,\text{-}2}$ $\ket{2,\text{-}2}$ & 0  & 2&in&  125.94 & 0.14  & -0.097 & -0.125  & -0.120  & present\\
     & -4 & $\ket{2,\text{-}2}$ $\ket{2,\text{-}2}$ & 0  & 2&in&  188.72 & 0.05  &  0.033 & -0.008  &  0.000  & \\
     & -4 & $\ket{2,\text{-}2}$ $\ket{2,\text{-}2}$ & 0  & 2&in&  227.71 & 0.60  & -0.150 & -0.170  & -0.130  & \\
\hline
40/40& -8 & $\ket{9/2,\text{-}9/2}$ $\ket{9/2,\text{-}7/2}$ & 0  & 0&el&  202.10 & 0.07  &  0.003 & -0.014  &  0.005  & \cite{Regal2003}\\
     & -7 & $\ket{9/2,\text{-}9/2}$ $\ket{9/2,\text{-}5/2}$ & 0  & 0&el&  224.21 & 0.05  & -0.001 & -0.018  &  0.003  & \\
     & -6 & $\ket{9/2,\text{-}7/2}$ $\ket{9/2,\text{-}5/2}$ & 0  & 0&el&  174.0  & 2.0   & -0.280 & -0.155  & -0.130  & \\
     & -7 & $\ket{9/2,\text{-}7/2}$ $\ket{9/2,\text{-}7/2}$ & 1  & 1&el&  198.81 & 0.05  & -0.010 & -0.044  & -0.024  & \cite{Ticknor2004}\\
     & -8 & $\ket{9/2,\text{-}7/2}$ $\ket{9/2,\text{-}7/2}$ & 1  & 1&el&  198.34 & 0.05  &  0.030 & -0.006  &  0.014  & \\
     &  7 & $\ket{9/2,7/2}$ $\ket{9/2,7/2}$ & 1  & 1&el&  436.3  & 0.5   & -0.484 & -0.444  & -0.440  & \cite{Regal2003}\\
\hline
39/41& -1 & $\ket{1,\text{-}0}$ $\ket{1,\text{-}1}$ & 0  & 0&el&  228.88 & 0.08  & -0.091 & -0.036  & -0.035  & \cite{PhysRevA.98.062712}\\
     & -2 & $\ket{1,\text{-}1}$ $\ket{1,\text{-}1}$ & 0  & 0&el&  149.84 & 0.06  &  0.100 &  0.083  &  0.082  & \\
     & -3 & $\ket{1,\text{-}1}$ $\ket{2,\text{-}2}$ & 0  & 0&el&  649.6  & 0.6   &  0.564 &  0.535  &  0.343  & \\
     &  3 & $\ket{1,1}$ $\ket{2,2}$ & 0  & 0&el&  341.5  & 0.20  & -0.018 &  0.097  &  0.101  & \\
     &  3 & $\ket{1,1}$ $\ket{2,2}$ & 0  & 0&el&  353.8  & 0.20  & -0.092 &  0.017  &  0.023  & \\
     &  2 & $\ket{1,1}$ $\ket{1,1}$ & 0  & 0&el&  139.27 & 0.04  & -0.072 & -0.020  & -0.016  & \\
     &  2 & $\ket{1,1}$ $\ket{1,1}$ & 0  & 0&el&  146.24 & 0.07  & -0.109 & -0.057  & -0.056  & \\
     &  2 & $\ket{1,1}$ $\ket{1,1}$ & 0  & 0&el&  338.12 & 0.07  & -0.057 &  0.032  &  0.024  & \\
     &  2 & $\ket{1,1}$ $\ket{1,1}$ & 0  & 0&el&  500.2  & 0.3   &  0.276 &  0.242  &  0.051  & \\
     &  2 & $\ket{1,1}$ $\ket{1,1}$ & 0  & 0&el&  518.4  & 0.1   &  0.072 &  0.049  & -0.124  & \\
     &  1 & $\ket{1,1}$ $\ket{1,0}$ & 0  & 0&el&   88.2  & 0.1   & -0.212 & -0.157  & -0.160  & \\
     &  1 & $\ket{1,1}$ $\ket{1,0}$ & 0  & 0&el&  160.05 & 0.06  & -0.020 &  0.034  &  0.037  & \\
     &  1 & $\ket{1,1}$ $\ket{1,0}$ & 0  & 0&el&  165.80 & 0.05  & -0.069 & -0.016  & -0.015  & \\
     &  1 & $\ket{1,1}$ $\ket{1,0}$ & 0  & 0&el&  344.4  & 0.1   & -0.006 &  0.073  &  0.053  & \\
     &  1 & $\ket{1,1}$ $\ket{1,0}$ & 0  & 0&el&  522.6  & 0.2   &  0.246 &  0.217  &  0.028  & \\
     &  1 & $\ket{1,1}$ $\ket{1,0}$ & 0  & 0&el&  553.1  & 0.1   &  0.243 &  0.229  &  0.069  & \\
     &  0 & $\ket{1,1}$ $\ket{1,\text{-}1}$ & 0  & 0&el&  189.88 & 0.05  & -0.053 &  0.000  &  0.001  & \\
     &  0 & $\ket{1,1}$ $\ket{1,\text{-}1}$ & 0  & 0&el&  348.4  & 0.1   &  0.040 &  0.110  &  0.076  & \\
     &  0 & $\ket{1,1}$ $\ket{1,\text{-}1}$ & 0  & 0&el&  384.91 & 0.07  & -0.044 &  0.049  &  0.049  & \\
     &  0 & $\ket{1,1}$ $\ket{1,\text{-}1}$ & 0  & 0&el&  553.5  & 0.2   &  0.249 &  0.223  &  0.044  & \\
     & -1 & $\ket{1,0}$ $\ket{1,\text{-}1}$ & 0  & 0&el&  228.88 & 0.08  & -0.091 & -0.036  & -0.035  & \\
     & -2 & $\ket{1,\text{-}1}$ $\ket{1,\text{-}1}$ & 0  & 0&el&  149.84 & 0.06  &  0.100 &  0.083  &  0.082  & \\
     & -3 & $\ket{1,\text{-}1}$ $\ket{2,\text{-}2}$ & 0  & 0&el&  649.6  & 0.6   &  0.564 &  0.535  &  0.343  & \\
\hline
41/41& -2 & $\ket{1,\text{-}1}$ $\ket{1,\text{-}1}$ & 0  & 0&el&   51.1  & 0.2   & -0.021 & -0.100  & -0.101  & \cite{PhysRevA.98.062712}\\
     & -1 & $\ket{1,\text{-}1}$ $\ket{1,0}$ & 0  & 0&el&   51.92 & 0.08  &  0.022 & -0.077  & -0.081  & \\
     &  2 & $\ket{1,1}$ $\ket{1,1}$ & 0  & 0&el&  409.2  & 0.2   & -0.160 &  0.053  &  0.068  & \cite{Chen2016}\\
     &  2 & $\ket{1,1}$ $\ket{1,1}$ & 0  & 0&el&  660.6  & 0.2   & -0.015 &  0.000  &  0.059  & \\
\hline
40/41&11/2 &$\ket{9/2,9/2}$ $\ket{1,1}$  & 0  & 2&el&  472.6  & 0.3   & -2.150 & -1.922  & -1.906  & \cite{PhysRevA.84.011601}\\ 
	   &11/2 &$\ket{9/2,9/2}$ $\ket{1,1}$  & 1  & 1&el&  432.9  & 0.3   & -2.651 & -2.280  & -2.252  & \\
	   &11/2 &$\ket{9/2,9/2}$ $\ket{1,1}$  & 0  & 0&el&  542.7  & 1.0   &  0.167 &  0.402  &  0.416  & \\ 
\hline
\end{tabular*}\\
\end{table*}
We base our analysis on Feshbach resonance data for the isotope combinations \K{39}-\K{39}~from D'Errico \textit{et al.} \cite{D_Errico_2007} and \cite{Fouche2019,PhysRevA.98.062712},  \K{40}-\K{40}~from Regal \textit{et al.} \cite{Regal2003,PhysRevLett.92.040403},  \K{39}-\K{41}~from Tanzi \textit{et al.} \cite{PhysRevA.98.062712},  \K{41}-\K{41}~ from Chen \textit{et al.} \cite{Chen2016} and Tanzi et al. \cite{PhysRevA.98.062712}, and  \K{40}-\K{41}~from Wu \textit{et al.} \cite{PhysRevA.84.011601}. We summarize the data  in Tab.\,\ref{tab:FR} with their quantum numbers and the reported experimental uncertainty. As quantum numbers we use the projection $M$ of the total angular momentum onto the field axis, the atom pair labels for dressed states  and the interval $l_\mathrm{min}-l_\mathrm{max}$ of the partial waves. The labels of the atomic dressed states are given by $\ket{f,m_f}$. The column `type' indicates, that for `el' the peak of the elastic part of the rate coefficient is taken and for `in' the sum of the inelastic contributions.\\
The evaluation uses atomic hyperfine and $g$ factors from \cite{Arimondo1977}. We fit the data in Tab.\,\ref{tab:FR} to the BO-potentials adjusting the branches at small ($R<R_\mathrm{in}$) and large ($R>R_\mathrm{out}$) internuclear separations. After few trials it became clear, that the 3 data points  for the \K{40}-\K{41}~ isotope combination show significant deviations (several times the experimental uncertainty) compared to all other isotope combinations. In the following, we thus exclude these resonances from the analysis and report only their resulting deviations in the final conclusion. Refitting all remaining resonances,  we obtain a normalized standard deviation of $\sigma=0.977$. These results are given in Tab.\,\ref{tab:FR} in column `o-c(1)' (observed-calculated) and in row 1 of Tab.\,\ref{Sigma} (labeled model (1)).\\
Analyzing the obtained fit for the different isotope combinations (see Tab. \,\ref{Sigma}, model (1)) reveals that the main part of the sum of squared weighted deviations stems from the isotope combination \K{39}-\K{41}, resulting in $\sigma=1.235$, whereas the other isotope combinations show values below 0.72. A separate fit to the data of the isotope combination \K{39}-\K{41}~only underlines the consistency of these observations  with a resulting normalized standard deviation of $\sigma=0.753$ (see Tab. \,\ref{Sigma}, model ($^{39}$K$^{41}$K)$_a$). We thus started an additional fit in an attempt to optimize the result for the  isotope combination \K{39}-\K{41} by applying the potentials from the separate fit of \K{39}-\K{41} above as initial values. Note that non-linear fits regularly give slightly different results depending on initial values since it is hard to  find the global minimum of the sum of squared weighted deviations. In this second fit we obtain almost the same overall fit quality with a normalized standard deviation of  $\sigma=0.993$ but now the main deviation lies in the \K{40}-\K{40}~ isotope combination with its individual value $\sigma=1.566$ as given in the third line of Tab.\,\ref{Sigma}, model ($^{39}$K$^{41}$K)$_b$). Since the reduced mas of $^{39}$K$^{41}$K and \K{40}-\K{40} are almost equal, the different behaviour of these two isotope combinations in the two different fits (model (1) and model ($^{39}$K$^{41}$K)$_b$) give a strong hint that mass  effects beyond the simple mass scaling of the rovibrational motion should be considered.
\begin{table}
 \caption{Overview of obtained standard deviations at different evaluation steps. The column `model' gives the same numbers as in Tab.\,\ref{tab:FR}. Columns labeled by isotope combinations show the contribution of that combination.}
\label{Sigma}
\begin{threeparttable}
\begin{tabular}{ c |c|c|c|c|c}
\multirow{2}{4em}{model} & \multicolumn{5}{c}{$\sigma$}\\    %
& ~\textbf{total}~ & \K{39}-\K{41}  & \K{40}-\K{40} &  \K{39}-\K{39}& \K{41}-\K{41}\\
\hline(1)& \textbf{0.977}  & 1.235 &0.476& 0.718&0.428\\
(39/41)$_\mathrm{a}$ & - & 0.753 & - & -&-\\
(39/41)$_b$& \textbf{0.993} &0.867& 1.566 & 0.991 &0.394\\
(2) & \textbf{0.786} & 0.952  & 0.541 & 0.593&0.556\\
(3) & \textbf{0.651} &0.737&0.428& 0.584 &0.607 \\
\end{tabular}
    \begin{tablenotes}
      \small
\item(1): BOA\\
\item(39/41)$_\mathrm{a}$: fit restricted to \K{39}-\K{41}\\
\item(39/41)$_\mathrm{b}$: as (1) using (39/41)$_\mathrm{a}$ as starting guess\\ 
\item(2): BOA with $BO_\mathrm{gen}$ corrections\\
\item(3): BOA with $BO_\mathrm{gen}$ and $BO_\mathrm{asym}$ corrections
    \end{tablenotes}
  \end{threeparttable}
\end{table}\\\newline
In the next step, we thus include beyond BO-corrections proportional to the reduced mass for the general case (i.e.  part $U_\mathrm{gen}$ from Eq.\,6). Because we  can only study the small variations between the naturally existing isotope combinations, it is of advantage to define one isotope combination as reference. This results in a parametrization of the full potentials with BO-corrections for molecule $AB$ given by
\begin{equation}
\label{adref}
\begin{split}
U(R)  = & U_\mathrm{BO}(R)+\\
 & BO_\mathrm{gen}(R)\left(1-\frac{\mu_\mathrm{ref}}{\mu_{AB}}\right)+\\
 & BO_\mathrm{asym}(R)\left(\frac{M_A-M_B}{M_A+M_B}\right)^2,
\end{split}
\end{equation}
where the factor of the electron mass in Eq.\,\ref{ad} is incorporated in the new functions $BO_\mathrm{gen}$ and $BO_\mathrm{asym}$ and $\mu_\mathrm{ref}$ is the reduced mass of the reference combination. Here we apply \K{39}-\K{39} as reference. For this combination we have a large number of $s$-wave resonances and additionally also $d$-wave resonances. Both together fix the asymptotic branch of the potentials. This is different for the  \K{39}-\K{41} isotope combination where only $s$-wave resonances have been measured. In principle,  $BO_\mathrm{gen}$ is a function of $R$, but the present data set is too small to derive such function from a fit with acceptable significance. Thus we simplify the condition by assuming correction functions to be proportional to the BO-potential and Eq.\,\ref{adref} reads now
\begin{equation}
\label{adref1}
\begin{split}
U(R)  = & U_\mathrm{BO}(R)\left(1+ BO_\mathrm{gen}\left(\vphantom{\left(\frac{M_A-M_B}{M_A+M_B}\right)^2}1-\frac{\mu_\mathrm{ref}}{\mu_{AB}}\right)\right.+\\
 & \left. BO_\mathrm{asym}\left(\frac{M_A-M_B}{M_A+M_B}\right)^2\right),
\end{split}
\end{equation}
where now $BO_\mathrm{gen}$ and $BO_\mathrm{asym}$ are fit parameters for the amplitude of the BO-corrections. A crude justification of this assumption is that the normal mass effect in atomic physics, e.g. the Rydberg constant and its nuclear mass dependence, show a similar form of the correction for the binding energy. Furthermore, a molecular potential describes the variation of the kinetic energy within the nuclear vibrational motion as function of $R$ and is therefore a measure of the coupling to the electron motion.\\
Starting with $BO_\mathrm{gen}$ for states \Xstate and \astate we perform a fit of all resonances adding the parameter for both electronic states and obtain a normalized standard deviation of $\sigma=0.786$ (the individual deviations are shown in column `o-c(2)' in Tab.\,\ref{tab:FR}). This value should be compared with the one from a fit of the pure BO-potentials  $\sigma=0.977$. Including beyond Born-Oppenheimer corrections apparently leads to a significantly better fit.\\
Looking again at details of the fit for the different isotope combinations in Tab.\,\ref{Sigma} model (2), we see that the combination \K{39}-\K{41} is described with $\sigma=0.952$ whereas the other three show values below 0.6. Since we removed the isotope combination \K{40}-\K{41} from the evaluation already earlier for other reason (see also \cite{Note11}), the former one is the only heteronuclear case remaining for which the standard deviation is significantly larger than the value seen in a separate fit (comp. above 0.753). Keeping in mind that the isotope combination with almost equal reduced mass, namely \K{40}-\K{40}, is well represented by the introduced BO-correction, we complement our model by  the heteronuclear extension of the BO-correction, which is already contained in Eq.\,\ref{adref1} by the parameter $BO_\mathrm{asym}$. The new fit results now in a standard deviation of $\sigma=0.651$ (see Tab.\,\ref{Sigma} model (3)) thus a further improvement compared to 0.786 from model (2). Additionally, all individual standard deviations are almost equal to the values obtained by separated fits.  The deviations of observation to calculation from the new fit are shown in column `o-c(3)' of Tab.\,\ref{tab:FR}. The sequential improvement of the fit quality including beyond Born-Oppenheimer terms underlines the significance of corrections beyond Born-Oppenheimer for the precise description of molecular potentials and the precision derivation of atomic scattering properties. \cite{MNote13}
\begin{table}
\fontsize{8pt}{13pt}\selectfont
\caption{Born-Oppenheimer corrections, according Eq.\,\ref{adref1}. The number indicates the model of evaluation: (2) adding BO-correction $BO_\mathrm{gen}$ for the general case, (3) adding BO-corrections  $BO_\mathrm{gen}$ and $BO_\mathrm{asym}$ for the general and heteronuclear case. Values in brackets are not significantly determined and effectively zero.}
\label{BOC}
\begin{tabular*}{0.8\columnwidth} {@{\extracolsep{\fill}}l|cc}
\hline
parameter & \Xstate& \astate \\
\hline
$BO_\mathrm{gen}$~~(2)& (-0.00003)  & -0.00046  \\
\hline
$BO_\mathrm{gen}$~~(3)& (-0.00006)  & -0.00046  \\
$BO_\mathrm{asym}$~(3)& 0.0057  & (-0.000001)  \\
\hline
\end{tabular*}
\end{table}\\
In Tab.\,\ref{BOC} we give the magnitude of the BO-corrections for the two electronic states \Xstate and $a^3\Sigma_u^+$. The uncertainty of the significantly determined parameters are about 20\,\%. For giving a better insight into the BO-correction we calculate the highest vibrational levels with the correction and compare them with the level energy setting the correction to zero. For the heaviest isotope \K{41}-\K{41} and thus the largest difference to the reference isotope \K{39}-\K{39} we obtain for the level $v=27$, $N=0$ of the state \astate a difference of 220\,kHz and for the state \Xstate ($v=87$, $N=0$) it is effectively zero, because here the influence by BO-correction appears only for heteronuclear isotope combinations.

\section{Discussions}
\begin{table}
\fontsize{8pt}{13pt}\selectfont
\caption{Scattering lengths in units of $a_0$ of all natural isotope combinations of K derived from the different potential models.}
\label{tab:scatl}
\begin{tabular*}{0.87\columnwidth} {c|c|c|c|c}  
\hline
isotope &~(1) \Xstate~ &~(3) \Xstate~ &~(1) \astate~  &~(3) \astate~ \\
\hline
39/39  &138.801 &138.759(20)  &-33.376  &-33.413(25)  \\  
39/40  &-2.669  &-2.707(15)   &-2031  &-2026(10)  \\ 
39/41  &113.094 &113.036(20)  &176.600  &176.688(25)  \\  
40/40  &104.416 &104.410(20)  &169.204  &169.288(25)  \\  
40/41  &-54.447 &-54.479(25)  &97.139  &97.186(20)  \\  
41/41  &85.400  &85.409(18)   &60.266  &60.317(18)  \\  
\hline
\end{tabular*}
\end{table}

From the different models we calculated the scattering lengths for the pure singlet and triplet state. The results are summarized in Tab.\,\ref{tab:scatl} for the different isotope combinations using model (1) (BO-approximation) and (3) (including all beyond BO-corrections). Because we choose the isotope combination \K{39}-\K{39} as reference one might expect no difference for the resulting scattering lengths  for this isotope pair when using model (1) or (3) respectively.  However, we do observe corrections (see Tab. \ref{tab:scatl}). Equal values for \K{39}-\K{39} would result if the evaluation in case (3) would only vary the BO-correction parameters and anything else be kept constant. But this will be not the optimal fit strategy, because in case (1), i.e. no BO-corrections, existing significant BO-corrections are distributed over the deviations of the fit over all isotopes and thus also the reference isotope is influenced. One can see such different distribution from the standard deviations of \K{39}-\K{39} given for model (1) and (3) in Tab.\,\ref{Sigma}, the former one is larger than the latter one. Because of this influence, we only give error estimates for the complete model including BO-corrections in Tab.\,\ref{tab:scatl} and the differences between model (1) and (3) do not show the true magnitude of the BO-correction. See also the calculation of the energy shift by the BO-correction as given at the end of section \ref{fit}.\\
A complete list of scattering lengths was reported in \cite{PhysRevA.78.012503}. The new values show a significant improvement by roughly a factor 5 of the error limit. The values agree in most cases within uncertainty limits despite the fact that the former evaluation could only incorporate Feshbach resonances for \K{39}-\K{39} and \K{40}-\K{40}. The paper stated that a weak indication of BO-corrections could be obtained from the resonances. But we believe the present evaluation shows this clearly. Additionally, we were able to study the difference between the homonuclear and heteronuclear case resulting in the values of $BO_\mathrm{gen}$ and $BO_\mathrm{asym}$.\\\newline
We obtained a significant contribution for the triplet state $a^3\Sigma^+_u$ by $BO_\mathrm{gen}$ for both fit cases but for the singlet state \Xstate only for the heteronuclear isotope pairs. This is probably related to the fact that the closest singlet state, namely $A^1\Sigma^+_u$, has $u$ symmetry compared to $g$ symmetry for the singlet ground state. These two can only couple by the symmetry breaking part of the Hamiltonian responsible for the BO-correction \cite{Bunker1973}. The situation for the triplet state is different, where the energetically closest is $b^3\Pi_u$ and has $u$ symmetry as the triplet ground state. We should note that the magnitudes of both effects, $BO_\mathrm{gen}$ and $BO_\mathrm{asym}$ cannot be directly compared, because the former one is referenced to \K{39}-\K{39} and thus describes only the difference between the isotope pairs whereas the latter indicates the total effect.\\\newline
We evaluated the isotope dependence by using the precise Feshbach spectroscopy and checked finally that the obtained BO-corrections have only little influence in the deep rovibrational levels measured by molecular spectroscopy, e.g. in Ref.\,\cite{PhysRevA.78.012503,Amiot1995,Pashov2008}, which have an uncertainty in the order of few thousands of cm$^{-1}$ or about 100\,MHz compared to 1\,MHz or better for the Feshbach spectroscopy. For this purpose we went back to the full data set from the spectroscopy for iterating the fit for obtaining the consistent description of the complete data set from molecular and Feshbach spectroscopy. The final parameter sets of the potentials are given in the appendix. 

\section{Conclusion and Outlook}
We use an analysis of the complete set of all known Feshbach resonances in different K isotope combinations to derive  potential energy curves for states \Xstate and \astate and find clear signatures of beyond BO-corrections. We base our work on the discussion of H$_2$ and HD molecules by van Vleck \cite{vanVleck1936} and find correction terms for the homonuclear and heteronuclear cases  when analysing homo- and heteronuclear isotope combinations of K respectively.  Unfortunately, our analysis of heteronuclear cases is restricted to the  \K{39}-\K{41} isotope combination, although, in principle, more isotope combinations exist. However, available Feshbach resonance data of the  \K{40}-\K{41} \cite{PhysRevA.84.011601} isotope  combination show very large deviations which are beyond a realistic description \cite{MNote35}. We therefore excluded this isotope combination from the analysis given in Sec.\,\ref{fit}. To allow for an extended analysis of heteronuclear beyond BO-corrections, it would be very much desirable to revisit observed Feshbach resonances in the $\ket{9/2,9/2}+\ket{1,1}$ channel of the \K{40}-\K{41} isotope combination  and extend measurements to resonances within other collision channels such as $\ket{9/2,\text{-}9/2}$+$\ket{1,1}$. In the same context, the \K{39}-\K{40} isotope is of great interest. Here, it would be particularly favorable to study collisions in the  $\ket{9/2,\text{-}7/2}$+$\ket{1,1}$ and $\ket{9/2,\text{-}5/2}$+$\ket{1,1}$ channels. In these channels well-separated Feshbach resonances in a magnetic field region below 200\,G should be found whereas sharp resonances in the $\ket{9/2,\text{-}9/2}$+$\ket{1,1}$ channel will be  overlapped by a very broad resonance. Furthermore, the above-mentioned channels will show sharp Feshbach resonances in the range of 800 to 850\,G. We believe that such studies will settle the discussion of the importance of BO-corrections in cases of homo- and hetero-nuclear pairs of homo-polar molecules.\\
In the same spirit, it would be very interesting to analyze Feshbach resonances in the different isotope combination of the homo-polar molecule Li$_2$. There exists a detailed analysis \cite{LeRoy2009} of  spectroscopic data of the \Xstate-$A^1\Sigma^+_u$ transition in the Li$_2$ considering homonuclear BO-corrections ($BO_\mathrm{gen}(R)$ from Eq.\,\ref{adref}). The study includes data from $^7$Li-$^6$Li isotopologue, however, the authors do not  mention any  need to distinguish between homo- and hetero-nuclear corrections. The data set for the $^7$Li-$^6$Li molecule is small compared to that of both homonuclear molecules $^7$Li-$^7$Li and $^6$Li-$^6$Li, thus it could be not sufficiently significant for the above mentioned distinction. For Li$_2$ there exist also measurements of Feshbach resonances for the homonuclear cases, see the latest report by Gerken et al. \cite{Gerken2019}, but nothing on $^7$Li-$^6$Li. Thus studies of Feshbach resonances of Li-Li would be very worth to investigate both homonuclear and heteronuclear beyond BO corrections.\\
We conclude that very interesting Feshbach spectroscopy is ahead of us to work out and highlight the importance of BO-corrections in the understanding of cold collisions.

\section{Acknowledgement}
We gratefully acknowledge financial support from the European Research Council through ERC Starting Grant POLAR and from the Deutsche Forschungsgemeinschaft (DFG) through CRC 1227 (DQ-mat), project A03 and FOR2247, project E5. K.K.V. and P.G. thank the Deutsche Forschungsgemeinschaft for financial support through Research Training Group 1991.

\section{Appendix}
Tab.\,\ref{tab:X}, \ref{tab:a} and \ref{tab:lr} show the potential parameters (defined in Eqs.\,\ref{uanal},  \ref{rep} and \ref{lrexp}) for the two states \Xstate and $a^3\Sigma_u^+$, as derived during the evaluation. These results are improved potentials compared to the published ones \cite{PhysRevA.78.012503}, not only because the Feshbach data were largely extended but also the number of potential parameters is significantly reduced leading to a more stringent potential form with less danger of showing tiny oscillatory unphysical effects.

\newpage
\begin{table}[ht]
\caption{Parameters of the analytic representation of the \Xstate state potential with adiabatic Born-Oppenheimer correction and reference isotopologue \K{39}-\K{39}. The energy reference is the dissociation asymptote. Parameters with $^\ast$ are set for continuous extrapolation of the potential.}
\label{tab:X}
\begin{tabular*}{0.8\columnwidth}{@{\extracolsep{\fill}}|cr|}
\hline
   \multicolumn{2}{|c|}{$R < R_\mathrm{in}=$ 2.87\,\AA}    \\
\hline
   $A^\ast$ & $-0.2600158561\times 10^{4}$\,\wn \\
   $B^\ast$ & $ 0.8053173040\times 10^{9}$ \,\wn\,\AA $^{12}$ \\
	$N_s$ & 12\\
\hline
   \multicolumn{2}{|c|}{$R_\mathrm{in} \leq R \leq R_\mathrm{out}=$ 12.000\,\AA} \\
\hline
    $b$ &   $-0.39$ \\
    $R_\mathrm{m}$ & 3.9243617\,\AA  \\
    $a_{0}$ &  $-4450.9007703$\,\wn\\
    $a_{1}$ &   0.159877863995326747              \,\wn\\
    $a_{2}$ & $ 0.141337574101676037\times 10^{5}$\,\wn\\
    $a_{3}$ & $ 0.107669620493846905\times 10^{5}$\,\wn\\
    $a_{4}$ & $-0.331314023322698995\times 10^{4}$\,\wn\\
    $a_{5}$ & $-0.163943210499613087\times 10^{5}$\,\wn\\
    $a_{6}$ & $-0.216334200177141829\times 10^{5}$\,\wn\\
    $a_{7}$ & $-0.384655804768731250\times 10^{5}$\,\wn\\
    $a_{8}$ & $-0.768229889574501722\times 10^{5}$\,\wn\\
    $a_{9}$ & $ 0.157896664088991121\times 10^{6}$\,\wn\\
   $a_{10}$ & $ 0.833691806464401074\times 10^{6}$\,\wn\\
   $a_{11}$ & $-0.115890452663354226\times 10^{7}$\,\wn\\
   $a_{12}$ & $-0.653607110081680864\times 10^{7}$\,\wn\\
   $a_{13}$ & $ 0.487172809603480622\times 10^{7}$\,\wn\\
   $a_{14}$ & $ 0.308101362964722812\times 10^{8}$\,\wn\\
   $a_{15}$ & $-0.863340173933527432\times 10^{7}$\,\wn\\
   $a_{16}$ & $-0.811804637748816609\times 10^{8}$\,\wn\\
   $a_{17}$ & $ 0.492251670364311151\times 10^{7}$\,\wn\\
   $a_{18}$ & $ 0.121156746090629265\times 10^{9}$\,\wn\\
   $a_{19}$ & $ 0.280059277888290165\times 10^{7}$\,\wn\\
   $a_{20}$ & $-0.968951931944736689\times 10^{8}$\,\wn\\
   $a_{21}$ & $-0.314874358611015789\times 10^{7}$\,\wn\\
   $a_{22}$ & $ 0.324661526246530302\times 10^{8}$\,\wn\\
\hline          
    $BO_\mathrm{gen}$  & (-0.00006) \\
    $BO_\mathrm{asym}$ &  0.00566 \\
\hline
\end{tabular*}
\end{table}

\begin{table}[h]

\caption{Parameters of the long range part of the potentials for both states \Xstate and $a^3\Sigma^+_u$. }
\label{tab:lr}
\begin{tabular*}{0.8\columnwidth}{@{\extracolsep{\fill}}|cr|}
\hline
   \multicolumn{2}{|c|}{$R_\mathrm{out} < R$}\\
\hline
  ${U_\infty}$ & 0.0\,\wn    \\
 ${C_6}$ &    0.1892338370$\times 10^{8}$\,\wn\AA$^6$      \\
 ${C_{8}}$ &  0.5706799528$\times 10^{9}$\,\wn\AA$^8$   \\
 ${C_{10}}$ & 0.1853042723$\times 10^{11}$\,\wn\AA$^{10}$   \\
 ${A_{\rm ex}}$ &$0.90092159\times 10^{4}$\,\wn\AA$^{-\gamma}$   \\
 ${\gamma}$ & 5.19500    \\
 ${\beta}$ & 2.13539\,\AA$^{-1}$   \\
\hline
\end{tabular*}
\end{table}

\begin{table}

\caption{Parameters of the analytic representation of the \astate state potential with adiabatic Born-Oppenheimer correction and reference isotopologue $^{39}$K$_2$. The energy reference is the dissociation asymptote. Parameters with $^\ast$ are set for continuous extrapolation of the potential.  }
\label{tab:a}
\begin{tabular*}{0.8\columnwidth}{@{\extracolsep{\fill}}|cr|}
\hline
\multicolumn{2}{|c|}{$R < R_\mathrm{in}=$ 4.755\,\AA}    \\
\hline
   $A^\ast$ & $-0.7009379657\times 10^{3}$\,\wn \\
   $B^\ast$ & $0.80690073665\times 10^{7}$ \,\wn\,\AA $^{6}$ \\
	$N_s$ & 6\\
\hline
   \multicolumn{2}{|c|}{$R_\mathrm{inn} \leq R \leq R_\mathrm{out}=$ 12.000\,\AA}    \\
\hline
    $b$ &   $-0.40$              \\
    $R_\mathrm{m}$ & 5.7347289\,\AA  \\
    $a_{0}$ &  $-255.0214692$\,\wn\\
    $a_{1}$ & $-0.013405598929310479$\,\wn\\
    $a_{2}$ & $ 0.153940442323125171\times 10^{4}$\,\wn\\
    $a_{3}$ & $-0.626944977828736569\times 10^{3}$\,\wn\\
    $a_{4}$ & $-0.147039918194012284\times 10^{4}$\,\wn\\
    $a_{5}$ & $ 0.238628331428504282\times 10^{3}$\,\wn\\
    $a_{6}$ & $-0.121465057044283844\times 10^{4}$\,\wn\\
    $a_{7}$ & $-0.131024472517054273\times 10^{5}$\,\wn\\
    $a_{8}$ & $ 0.410390478256789502\times 10^{5}$\,\wn\\
    $a_{9}$ & $ 0.585609645570106004\times 10^{5}$\,\wn\\
   $a_{10}$ & $-0.316660644987405278\times 10^{6}$\,\wn\\
   $a_{11}$ & $ 0.178579875710784399\times 10^{6}$\,\wn\\
   $a_{12}$ & $ 0.690085326716458891\times 10^{6}$\,\wn\\
   $a_{13}$ & $-0.116538893384502688\times 10^{7}$\,\wn\\
   $a_{14}$ & $ 0.541518493723396794\times 10^{6}$\,\wn\\
\hline
    $BO_\mathrm{gen}$  & -0.000465 \\
    $BO_\mathrm{asym}$ & (-0.000001) \\
\hline
\end{tabular*}
\end{table}


\newpage
\begin{thebibliography}{35}
\expandafter\ifx\csname natexlab\endcsname\relax\def\natexlab#1{#1}\fi
\expandafter\ifx\csname bibnamefont\endcsname\relax
  \def\bibnamefont#1{#1}\fi
\expandafter\ifx\csname bibfnamefont\endcsname\relax
  \def\bibfnamefont#1{#1}\fi
\expandafter\ifx\csname citenamefont\endcsname\relax
  \def\citenamefont#1{#1}\fi
\expandafter\ifx\csname url\endcsname\relax
  \def\url#1{\texttt{#1}}\fi
\expandafter\ifx\csname urlprefix\endcsname\relax\def\urlprefix{URL }\fi
\providecommand{\bibinfo}[2]{#2}
\providecommand{\eprint}[2][]{\url{#2}}

\bibitem[{\citenamefont{Born and Oppenheimer}(1927)}]{BornOpp}
\bibinfo{author}{\bibfnamefont{M.}~\bibnamefont{Born}} \bibnamefont{and}
  \bibinfo{author}{\bibfnamefont{R.}~\bibnamefont{Oppenheimer}},
  \bibinfo{journal}{Annalen der Physik} \textbf{\bibinfo{volume}{389}},
  \bibinfo{pages}{457} (\bibinfo{year}{1927}).

\bibitem[{\citenamefont{Herman and Asgharian}(1966)}]{HERMAN1966305}
\bibinfo{author}{\bibfnamefont{R.}~\bibnamefont{Herman}} \bibnamefont{and}
  \bibinfo{author}{\bibfnamefont{A.}~\bibnamefont{Asgharian}},
  \bibinfo{journal}{Journal of Molecular Spectroscopy}
  \textbf{\bibinfo{volume}{19}}, \bibinfo{pages}{305 } (\bibinfo{year}{1966}),
  ISSN \bibinfo{issn}{0022-2852}.

\bibitem[{\citenamefont{Watson}(1980)}]{Watson1980}
\bibinfo{author}{\bibfnamefont{J.~K.~G.} \bibnamefont{Watson}},
  \bibinfo{journal}{J. MOL. SPECTROSC. 80, 411}  (\bibinfo{year}{1980}).

\bibitem[{\citenamefont{Bunker}(1977)}]{Bunker1977}
\bibinfo{author}{\bibfnamefont{P.~R.} \bibnamefont{Bunker}},
  \bibinfo{journal}{J. MOL. SPECTROSC. 68, 367}  (\bibinfo{year}{1977}).

\bibitem[{\citenamefont{Che et~al.}(2007)\citenamefont{Che, Ren, Wang, Dong,
  Dai, Wang, Zhang, Yang, Sheng, Li et~al.}}]{Che1061}
\bibinfo{author}{\bibfnamefont{L.}~\bibnamefont{Che}},
  \bibinfo{author}{\bibfnamefont{Z.}~\bibnamefont{Ren}},
  \bibinfo{author}{\bibfnamefont{X.}~\bibnamefont{Wang}},
  \bibinfo{author}{\bibfnamefont{W.}~\bibnamefont{Dong}},
  \bibinfo{author}{\bibfnamefont{D.}~\bibnamefont{Dai}},
  \bibinfo{author}{\bibfnamefont{X.}~\bibnamefont{Wang}},
  \bibinfo{author}{\bibfnamefont{D.~H.} \bibnamefont{Zhang}},
  \bibinfo{author}{\bibfnamefont{X.}~\bibnamefont{Yang}},
  \bibinfo{author}{\bibfnamefont{L.}~\bibnamefont{Sheng}},
  \bibinfo{author}{\bibfnamefont{G.}~\bibnamefont{Li}}, \bibnamefont{et~al.},
  \bibinfo{journal}{Science} \textbf{\bibinfo{volume}{317}},
  \bibinfo{pages}{1061} (\bibinfo{year}{2007}).

\bibitem[{\citenamefont{Kn{\"o}ckel et~al.}(2004)\citenamefont{Kn{\"o}ckel,
  Bodermann, and Tiemann}}]{Knoeckel2004}
\bibinfo{author}{\bibfnamefont{H.}~\bibnamefont{Kn{\"o}ckel}},
  \bibinfo{author}{\bibfnamefont{B.}~\bibnamefont{Bodermann}},
  \bibnamefont{and} \bibinfo{author}{\bibfnamefont{E.}~\bibnamefont{Tiemann}},
  \bibinfo{journal}{The European Physical Journal D - Atomic, Molecular,
  Optical and Plasma Physics} \textbf{\bibinfo{volume}{28}},
  \bibinfo{pages}{199} (\bibinfo{year}{2004}), ISSN \bibinfo{issn}{1434-6079}.

\bibitem[{\citenamefont{Falke et~al.}(2008)\citenamefont{Falke, Kn\"ockel,
  Friebe, Riedmann, Tiemann, and Lisdat}}]{PhysRevA.78.012503}
\bibinfo{author}{\bibfnamefont{S.}~\bibnamefont{Falke}},
  \bibinfo{author}{\bibfnamefont{H.}~\bibnamefont{Kn\"ockel}},
  \bibinfo{author}{\bibfnamefont{J.}~\bibnamefont{Friebe}},
  \bibinfo{author}{\bibfnamefont{M.}~\bibnamefont{Riedmann}},
  \bibinfo{author}{\bibfnamefont{E.}~\bibnamefont{Tiemann}}, \bibnamefont{and}
  \bibinfo{author}{\bibfnamefont{C.}~\bibnamefont{Lisdat}},
  \bibinfo{journal}{Phys. Rev. A} \textbf{\bibinfo{volume}{78}},
  \bibinfo{pages}{012503} (\bibinfo{year}{2008}).

\bibitem[{\citenamefont{F\`arn\`ik et~al.}(2002)\citenamefont{F\`arn\`ik,
  Davis, A.~Kostin, Polyansky, Tennyson, and Nesbitt}}]{Farnik2016}
\bibinfo{author}{\bibfnamefont{M.}~\bibnamefont{F\`arn\`ik}},
  \bibinfo{author}{\bibfnamefont{S.}~\bibnamefont{Davis}},
  \bibinfo{author}{\bibfnamefont{M.}~\bibnamefont{A.~Kostin}},
  \bibinfo{author}{\bibfnamefont{O.}~\bibnamefont{Polyansky}},
  \bibinfo{author}{\bibfnamefont{J.}~\bibnamefont{Tennyson}}, \bibnamefont{and}
  \bibinfo{author}{\bibfnamefont{D.}~\bibnamefont{Nesbitt}},
  \bibinfo{journal}{The Journal of Chemical Physics}
  \textbf{\bibinfo{volume}{116}}, \bibinfo{pages}{6146} (\bibinfo{year}{2002}).

\bibitem[{\citenamefont{Chin et~al.}(2010{\natexlab{a}})\citenamefont{Chin,
  Grimm, Julienne, and Tiesinga}}]{RevModPhys.82.1225}
\bibinfo{author}{\bibfnamefont{C.}~\bibnamefont{Chin}},
  \bibinfo{author}{\bibfnamefont{R.}~\bibnamefont{Grimm}},
  \bibinfo{author}{\bibfnamefont{P.}~\bibnamefont{Julienne}}, \bibnamefont{and}
  \bibinfo{author}{\bibfnamefont{E.}~\bibnamefont{Tiesinga}},
  \bibinfo{journal}{Rev. Mod. Phys.} \textbf{\bibinfo{volume}{82}},
  \bibinfo{pages}{1225} (\bibinfo{year}{2010}{\natexlab{a}}).

\bibitem[{\citenamefont{Regal et~al.}(2003)\citenamefont{Regal, Ticknor, Bohn,
  and Jin}}]{PhysRevLett.90.053201}
\bibinfo{author}{\bibfnamefont{C.~A.} \bibnamefont{Regal}},
  \bibinfo{author}{\bibfnamefont{C.}~\bibnamefont{Ticknor}},
  \bibinfo{author}{\bibfnamefont{J.~L.} \bibnamefont{Bohn}}, \bibnamefont{and}
  \bibinfo{author}{\bibfnamefont{D.~S.} \bibnamefont{Jin}},
  \bibinfo{journal}{Phys. Rev. Lett.} \textbf{\bibinfo{volume}{90}},
  \bibinfo{pages}{053201} (\bibinfo{year}{2003}).

\bibitem[{\citenamefont{D{\textquotesingle}Errico
  et~al.}(2007)\citenamefont{D{\textquotesingle}Errico, Zaccanti, Fattori,
  Roati, Inguscio, Modugno, and Simoni}}]{D_Errico_2007}
\bibinfo{author}{\bibfnamefont{C.}~\bibnamefont{D{\textquotesingle}Errico}},
  \bibinfo{author}{\bibfnamefont{M.}~\bibnamefont{Zaccanti}},
  \bibinfo{author}{\bibfnamefont{M.}~\bibnamefont{Fattori}},
  \bibinfo{author}{\bibfnamefont{G.}~\bibnamefont{Roati}},
  \bibinfo{author}{\bibfnamefont{M.}~\bibnamefont{Inguscio}},
  \bibinfo{author}{\bibfnamefont{G.}~\bibnamefont{Modugno}}, \bibnamefont{and}
  \bibinfo{author}{\bibfnamefont{A.}~\bibnamefont{Simoni}},
  \bibinfo{journal}{New Journal of Physics} \textbf{\bibinfo{volume}{9}},
  \bibinfo{pages}{223} (\bibinfo{year}{2007}).

\bibitem[{\citenamefont{Wu et~al.}(2011)\citenamefont{Wu, Santiago, Park,
  Ahmadi, and Zwierlein}}]{PhysRevA.84.011601}
\bibinfo{author}{\bibfnamefont{C.-H.} \bibnamefont{Wu}},
  \bibinfo{author}{\bibfnamefont{I.}~\bibnamefont{Santiago}},
  \bibinfo{author}{\bibfnamefont{J.~W.} \bibnamefont{Park}},
  \bibinfo{author}{\bibfnamefont{P.}~\bibnamefont{Ahmadi}}, \bibnamefont{and}
  \bibinfo{author}{\bibfnamefont{M.~W.} \bibnamefont{Zwierlein}},
  \bibinfo{journal}{Phys. Rev. A} \textbf{\bibinfo{volume}{84}},
  \bibinfo{pages}{011601} (\bibinfo{year}{2011}).

\bibitem[{\citenamefont{Chen et~al.}(2016{\natexlab{a}})\citenamefont{Chen,
  Yao, Wu, Liu, Wang, Wang, Chen, and Pan}}]{PhysRevA.94.033408}
\bibinfo{author}{\bibfnamefont{H.-Z.} \bibnamefont{Chen}},
  \bibinfo{author}{\bibfnamefont{X.-C.} \bibnamefont{Yao}},
  \bibinfo{author}{\bibfnamefont{Y.-P.} \bibnamefont{Wu}},
  \bibinfo{author}{\bibfnamefont{X.-P.} \bibnamefont{Liu}},
  \bibinfo{author}{\bibfnamefont{X.-Q.} \bibnamefont{Wang}},
  \bibinfo{author}{\bibfnamefont{Y.-X.} \bibnamefont{Wang}},
  \bibinfo{author}{\bibfnamefont{Y.-A.} \bibnamefont{Chen}}, \bibnamefont{and}
  \bibinfo{author}{\bibfnamefont{J.-W.} \bibnamefont{Pan}},
  \bibinfo{journal}{Phys. Rev. A} \textbf{\bibinfo{volume}{94}},
  \bibinfo{pages}{033408} (\bibinfo{year}{2016}{\natexlab{a}}).

\bibitem[{\citenamefont{Tanzi et~al.}(2018)\citenamefont{Tanzi, Cabrera, Sanz,
  Cheiney, Tomza, and Tarruell}}]{PhysRevA.98.062712}
\bibinfo{author}{\bibfnamefont{L.}~\bibnamefont{Tanzi}},
  \bibinfo{author}{\bibfnamefont{C.~R.} \bibnamefont{Cabrera}},
  \bibinfo{author}{\bibfnamefont{J.}~\bibnamefont{Sanz}},
  \bibinfo{author}{\bibfnamefont{P.}~\bibnamefont{Cheiney}},
  \bibinfo{author}{\bibfnamefont{M.}~\bibnamefont{Tomza}}, \bibnamefont{and}
  \bibinfo{author}{\bibfnamefont{L.}~\bibnamefont{Tarruell}},
  \bibinfo{journal}{Phys. Rev. A} \textbf{\bibinfo{volume}{98}},
  \bibinfo{pages}{062712} (\bibinfo{year}{2018}).

\bibitem[{\citenamefont{Chin et~al.}(2010{\natexlab{b}})\citenamefont{Chin,
  Grimm, Julienne, and Tiesinga}}]{RevModPhysFesh}
\bibinfo{author}{\bibfnamefont{C.}~\bibnamefont{Chin}},
  \bibinfo{author}{\bibfnamefont{R.}~\bibnamefont{Grimm}},
  \bibinfo{author}{\bibfnamefont{P.}~\bibnamefont{Julienne}}, \bibnamefont{and}
  \bibinfo{author}{\bibfnamefont{E.}~\bibnamefont{Tiesinga}},
  \bibinfo{journal}{Rev. Mod. Phys.} \textbf{\bibinfo{volume}{82}},
  \bibinfo{pages}{1225} (\bibinfo{year}{2010}{\natexlab{b}}).

\bibitem[{\citenamefont{van Vleck}(1936)}]{vanVleck1936}
\bibinfo{author}{\bibfnamefont{J.~H.} \bibnamefont{van Vleck}},
  \bibinfo{journal}{J. Chem.Phys. 4, 327}  (\bibinfo{year}{1936}).

\bibitem[{\citenamefont{Schulze et~al.}(2018)\citenamefont{Schulze, Hartmann,
  Voges, Gempel, Tiemann, Zenesini, and Ospelkaus}}]{SchulzeBEC2018}
\bibinfo{author}{\bibfnamefont{T.~A.} \bibnamefont{Schulze}},
  \bibinfo{author}{\bibfnamefont{T.}~\bibnamefont{Hartmann}},
  \bibinfo{author}{\bibfnamefont{K.~K.} \bibnamefont{Voges}},
  \bibinfo{author}{\bibfnamefont{M.~W.} \bibnamefont{Gempel}},
  \bibinfo{author}{\bibfnamefont{E.}~\bibnamefont{Tiemann}},
  \bibinfo{author}{\bibfnamefont{A.}~\bibnamefont{Zenesini}}, \bibnamefont{and}
  \bibinfo{author}{\bibfnamefont{S.}~\bibnamefont{Ospelkaus}},
  \bibinfo{journal}{Phys. Rev. A} \textbf{\bibinfo{volume}{97}},
  \bibinfo{pages}{023623} (\bibinfo{year}{2018}).

\bibitem[{\citenamefont{Hartmann et~al.}(2019)\citenamefont{Hartmann, Schulze,
  Voges, Gersema, Gempel, Tiemann, Zenesini, and Ospelkaus}}]{Hartmann2019}
\bibinfo{author}{\bibfnamefont{T.}~\bibnamefont{Hartmann}},
  \bibinfo{author}{\bibfnamefont{T.~A.} \bibnamefont{Schulze}},
  \bibinfo{author}{\bibfnamefont{K.~K.} \bibnamefont{Voges}},
  \bibinfo{author}{\bibfnamefont{P.}~\bibnamefont{Gersema}},
  \bibinfo{author}{\bibfnamefont{M.~W.} \bibnamefont{Gempel}},
  \bibinfo{author}{\bibfnamefont{E.}~\bibnamefont{Tiemann}},
  \bibinfo{author}{\bibfnamefont{A.}~\bibnamefont{Zenesini}}, \bibnamefont{and}
  \bibinfo{author}{\bibfnamefont{S.}~\bibnamefont{Ospelkaus}},
  \bibinfo{journal}{Phys. Rev. A} \textbf{\bibinfo{volume}{99}},
  \bibinfo{pages}{032711} (\bibinfo{year}{2019}).

\bibitem[{\citenamefont{Weber et~al.}(2003)\citenamefont{Weber, Herbig, Mark,
  N\"agerl, and Grimm}}]{PhysRevLett.91.123201}
\bibinfo{author}{\bibfnamefont{T.}~\bibnamefont{Weber}},
  \bibinfo{author}{\bibfnamefont{J.}~\bibnamefont{Herbig}},
  \bibinfo{author}{\bibfnamefont{M.}~\bibnamefont{Mark}},
  \bibinfo{author}{\bibfnamefont{H.-C.} \bibnamefont{N\"agerl}},
  \bibnamefont{and} \bibinfo{author}{\bibfnamefont{R.}~\bibnamefont{Grimm}},
  \bibinfo{journal}{Phys. Rev. Lett.} \textbf{\bibinfo{volume}{91}},
  \bibinfo{pages}{123201} (\bibinfo{year}{2003}).

\bibitem[{\citenamefont{AU~Gregory et~al.}(2019)\citenamefont{AU~Gregory, Frye,
  Blackmore, Bridge, Sawant, Hutson, and Cornish}}]{Gregory2019p}
\bibinfo{author}{\bibfnamefont{P.~D.} \bibnamefont{AU~Gregory}},
  \bibinfo{author}{\bibfnamefont{M.~D.} \bibnamefont{Frye}},
  \bibinfo{author}{\bibfnamefont{J.~A.} \bibnamefont{Blackmore}},
  \bibinfo{author}{\bibfnamefont{E.~M.} \bibnamefont{Bridge}},
  \bibinfo{author}{\bibfnamefont{R.}~\bibnamefont{Sawant}},
  \bibinfo{author}{\bibfnamefont{J.~M.} \bibnamefont{Hutson}},
  \bibnamefont{and} \bibinfo{author}{\bibfnamefont{S.~L.}
  \bibnamefont{Cornish}}, \bibinfo{journal}{Nature Communications} p.
  \bibinfo{pages}{3104} (\bibinfo{year}{2019}).

\bibitem[{\citenamefont{Fouch\'e et~al.}(2019)\citenamefont{Fouch\'e, Boiss\'e,
  Berthet, Lepoutre, Simoni, and Bourdel1}}]{Fouche2019}
\bibinfo{author}{\bibfnamefont{L.}~\bibnamefont{Fouch\'e}},
  \bibinfo{author}{\bibfnamefont{A.}~\bibnamefont{Boiss\'e}},
  \bibinfo{author}{\bibfnamefont{G.}~\bibnamefont{Berthet}},
  \bibinfo{author}{\bibfnamefont{S.}~\bibnamefont{Lepoutre}},
  \bibinfo{author}{\bibfnamefont{A.}~\bibnamefont{Simoni}}, \bibnamefont{and}
  \bibinfo{author}{\bibfnamefont{T.}~\bibnamefont{Bourdel1}},
  \bibinfo{journal}{Phys. Rev. A 99, 022701}  (\bibinfo{year}{2019}).

\bibitem[{\citenamefont{Regal and Jin}(2004)}]{Regal2003}
\bibinfo{author}{\bibfnamefont{C.~A.} \bibnamefont{Regal}} \bibnamefont{and}
  \bibinfo{author}{\bibfnamefont{D.~S.} \bibnamefont{Jin}},
  \bibinfo{journal}{Phys. Rev. Lett.} \textbf{\bibinfo{volume}{90}},
  \bibinfo{pages}{230404} (\bibinfo{year}{2004}).

\bibitem[{\citenamefont{Ticknor et~al.}(2004)\citenamefont{Ticknor, Regal, Jin,
  and Bohn}}]{Ticknor2004}
\bibinfo{author}{\bibfnamefont{C.}~\bibnamefont{Ticknor}},
  \bibinfo{author}{\bibfnamefont{C.~A.} \bibnamefont{Regal}},
  \bibinfo{author}{\bibfnamefont{D.~S.} \bibnamefont{Jin}}, \bibnamefont{and}
  \bibinfo{author}{\bibfnamefont{J.~L.} \bibnamefont{Bohn}},
  \bibinfo{journal}{Phys. Rev. A 69, 042712}  (\bibinfo{year}{2004}).

\bibitem[{\citenamefont{Chen et~al.}(2016{\natexlab{b}})\citenamefont{Chen,
  Yao, Wu, Liu, Wang, Wang, Chen, , and Pan}}]{Chen2016}
\bibinfo{author}{\bibfnamefont{H.-Z.} \bibnamefont{Chen}},
  \bibinfo{author}{\bibfnamefont{X.-C.} \bibnamefont{Yao}},
  \bibinfo{author}{\bibfnamefont{Y.-P.} \bibnamefont{Wu}},
  \bibinfo{author}{\bibfnamefont{X.-P.} \bibnamefont{Liu}},
  \bibinfo{author}{\bibfnamefont{X.-Q.} \bibnamefont{Wang}},
  \bibinfo{author}{\bibfnamefont{Y.-X.} \bibnamefont{Wang}},
  \bibinfo{author}{\bibfnamefont{Y.-A.} \bibnamefont{Chen}}, ,
  \bibnamefont{and} \bibinfo{author}{\bibfnamefont{J.-W.} \bibnamefont{Pan}},
  \bibinfo{journal}{Phys. Rev. A94, 033408}
  (\bibinfo{year}{2016}{\natexlab{b}}).

\bibitem[{\citenamefont{Regal et~al.}(2004)\citenamefont{Regal, Greiner, and
  Jin}}]{PhysRevLett.92.040403}
\bibinfo{author}{\bibfnamefont{C.~A.} \bibnamefont{Regal}},
  \bibinfo{author}{\bibfnamefont{M.}~\bibnamefont{Greiner}}, \bibnamefont{and}
  \bibinfo{author}{\bibfnamefont{D.~S.} \bibnamefont{Jin}},
  \bibinfo{journal}{Phys. Rev. Lett.} \textbf{\bibinfo{volume}{92}},
  \bibinfo{pages}{040403} (\bibinfo{year}{2004}).

\bibitem[{\citenamefont{Arimondo et~al.}(1977)\citenamefont{Arimondo, Inguscio,
  and Violino}}]{Arimondo1977}
\bibinfo{author}{\bibfnamefont{E.}~\bibnamefont{Arimondo}},
  \bibinfo{author}{\bibfnamefont{M.}~\bibnamefont{Inguscio}}, \bibnamefont{and}
  \bibinfo{author}{\bibfnamefont{P.}~\bibnamefont{Violino}},
  \bibinfo{journal}{Rev. Mod. Phys. 49, 31}  (\bibinfo{year}{1977}).

\bibitem[{Not()}]{Note11}
\bibinfo{note}{At the bottom of Tab.\,\ref{tab:FR} we give the deviations for
  the removed isotope combination derived from the three fit models. The
  magnitude of the deviations does not show a significant difference between
  the three models described, thus it is independent of type of modeling and
  not related to BO-corrections. A reobservation of the Feshbach resonances of
  $^{40}\textrm{K}$-$^{41}\textrm{K}$ would be very much desirable.}

\bibitem[{MNo({\natexlab{a}})}]{MNote13}
\bibinfo{note}{Note, that within our many trials of non-linear fitting we found
  one solution excluding BO-corrections in which the
  $\sigma($\K{39}-\K{39}$)=0.924$ and $\sigma($\K{39}-\K{41}$)=0.988$ from the
  main body of data are almost equally distributed. But both are significantly
  larger than the ones derived from separate fits of the isotope combinations,
  namely 0.593 and 0.753. Thus also this fit shows the need of BO-corrections.}

\bibitem[{\citenamefont{Bunker}(1973)}]{Bunker1973}
\bibinfo{author}{\bibfnamefont{P.~R.} \bibnamefont{Bunker}},
  \bibinfo{journal}{J. Mol. Spectrosc. 46, 119}  (\bibinfo{year}{1973}).

\bibitem[{\citenamefont{C.Amiot et~al.}(1995)\citenamefont{C.Amiot, Verg{\`e}s,
  and Fellows}}]{Amiot1995}
\bibinfo{author}{\bibnamefont{C.Amiot}},
  \bibinfo{author}{\bibfnamefont{J.}~\bibnamefont{Verg{\`e}s}},
  \bibnamefont{and} \bibinfo{author}{\bibfnamefont{C.}~\bibnamefont{Fellows}},
  \bibinfo{journal}{J. Chem. Phys. 103, 3350}  (\bibinfo{year}{1995}).

\bibitem[{\citenamefont{Pashov et~al.}(2008)\citenamefont{Pashov, Popov,
  Kn\"ockel, and Tiemann}}]{Pashov2008}
\bibinfo{author}{\bibfnamefont{A.}~\bibnamefont{Pashov}},
  \bibinfo{author}{\bibfnamefont{P.}~\bibnamefont{Popov}},
  \bibinfo{author}{\bibfnamefont{H.}~\bibnamefont{Kn\"ockel}},
  \bibnamefont{and} \bibinfo{author}{\bibfnamefont{E.}~\bibnamefont{Tiemann}},
  \bibinfo{journal}{Eur. Phys. J. D 46, 241}  (\bibinfo{year}{2008}).

\bibitem[{MNo({\natexlab{b}})}]{MNote35}
\bibinfo{note}{By sharing our manuscript with M. Zwierlein \cite{MNote34}, we
  learned that for the two resonance at lower fields the authors extrapolated
  the field values from low field calibrations and therefore an error of 2 G is
  not unlikely. For the broad resonance at 542.7 G an uncertainty of 0.5 G
  would be a good estimate because an improved calibration was applied.}

\bibitem[{\citenamefont{LeRoy et~al.}(2009)\citenamefont{LeRoy, Dattani, Coxon,
  Ross, Crozet, and Linton}}]{LeRoy2009}
\bibinfo{author}{\bibfnamefont{R.~J.} \bibnamefont{LeRoy}},
  \bibinfo{author}{\bibfnamefont{N.~S.} \bibnamefont{Dattani}},
  \bibinfo{author}{\bibfnamefont{J.~A.} \bibnamefont{Coxon}},
  \bibinfo{author}{\bibfnamefont{A.~J.} \bibnamefont{Ross}},
  \bibinfo{author}{\bibfnamefont{P.}~\bibnamefont{Crozet}}, \bibnamefont{and}
  \bibinfo{author}{\bibfnamefont{C.}~\bibnamefont{Linton}},
  \bibinfo{journal}{J. Chem. Phys. 131, 204309}  (\bibinfo{year}{2009}).

\bibitem[{\citenamefont{Gerken et~al.}(2019)\citenamefont{Gerken, Tran,
  H\"afner, Tiemann, Zhu, and Weidem\"uller}}]{Gerken2019}
\bibinfo{author}{\bibfnamefont{M.}~\bibnamefont{Gerken}},
  \bibinfo{author}{\bibfnamefont{B.}~\bibnamefont{Tran}},
  \bibinfo{author}{\bibfnamefont{S.}~\bibnamefont{H\"afner}},
  \bibinfo{author}{\bibfnamefont{E.}~\bibnamefont{Tiemann}},
  \bibinfo{author}{\bibfnamefont{B.}~\bibnamefont{Zhu}}, \bibnamefont{and}
  \bibinfo{author}{\bibfnamefont{M.}~\bibnamefont{Weidem\"uller}},
  \bibinfo{journal}{Phys. Rev. A} \textbf{\bibinfo{volume}{100}},
  \bibinfo{pages}{050701} (\bibinfo{year}{2019}).

\bibitem[{MNo({\natexlab{c}})}]{MNote34}
\bibinfo{note}{M. Zwierlein, private communication (2019)}.

\end{thebibliography}
\end{document}